\documentclass[12pt]{article}
\usepackage{epsfig}

\def\lsim{\raise0.3ex\hbox{$<$\kern-0.75em\raise-1.1ex\hbox{$\sim$}}}
\def\gsim{\raise0.3ex\hbox{$>$\kern-0.75em\raise-1.1ex\hbox{$\sim$}}}

\def\beq{\begin{equation}}
\def\eeq{\end{equation}}
\def\bea{\begin{eqnarray}}
\def\eea{\end{eqnarray}}
\def\noi{\noindent}
\begin{document}

\rightline{UCOFIS 1/02}
\rightline{June 2002}

\vspace{1cm}

\begin{center}
{\Large\bf A simple model for}\\ 
{\Large\bf nuclear structure functions at small $x$}\\
{\Large\bf in the dipole picture}
\vspace{1cm}
 
N. Armesto
 
\vspace{0.2cm}
 
{\it Departamento de F\'{\i}sica, M\'odulo C2, Planta baja, Campus
de Rabanales,}\\
{\it Universidad de C\'ordoba, E-14071 C\'ordoba, Spain} \\
 
\end{center}

\vspace{2cm}
{\small
A simple model for nuclear structure functions in the region of
small $x$ and small and
moderate $Q^2$,
is presented. It is a
parameter-free extension, in the Glauber-Gribov approach to nuclear collisions,
of a saturation model for the nucleon.
A reasonable agreement with experimental data
on ratios of nuclear structure functions
is obtained. Nuclear effects in the longitudinal-to-transverse cross section
ratios are found to be small.
Predictions of the model for values of $x$ smaller than those
available to present experiments are given. The unintegrated gluon
distribution and the behaviour of the saturation scale
which result from this model are shown and discussed.
}



\newpage

\section{Introduction} \label{intro}

The differences between the structure functions measured in nucleons and nuclei 
\cite{emc}, the so-called EMC effect, are a very important feature for the
study of nuclear structure and nuclear collisions. At small values of the
Bjorken variable $x$ ($\lsim \ 0.01$,
shadowing region), the
structure function $F_2$ per nucleon turns out to be smaller in nuclei than
in a free nucleon.
The nature of this shadowing is well understood
qualitatively:
In the rest frame of the nucleus,
the incoming photon splits, at high enough $Q^2$, into a $q\bar q$ pair long
before reaching the nucleus, and this $q\bar q$ pair interacts with it
with typical hadronic cross sections, which results in absorption
\cite{brodsky,nikolaev,nikolaev2,kope,noso}; in this way nuclear shadowing is a
consequence of
multiple scattering and is thus related with diffraction \cite{funnuc}.
An equivalent explanation in the
frame in which the nucleus is moving fast, is that gluon recombination due
to the overlap of the gluon clouds from different nucleons, makes gluon density
in nucleus with mass number
$A$ smaller than $A$ times that in a free nucleon \cite{glr,mq}.
These studies have received great theoretical impulse with the development of
semiclassical ideas in QCD and the appearance of non-linear equations for
evolution in $x$ in this
framework
\cite{semi,balit,mueller1,numer,levin,misha,misha2},
although saturation appears to be
different from shadowing \cite{mueller2} (see
\cite{carlinhos} for a simple geometrical approach in this framework).

On the other hand, a different approach is taken in \cite{gnucl}: parton
densities inside the nucleus are parametrized at some scale
$Q_0^2\sim 2$ GeV$^2$ and then evolved using the DGLAP \cite{dglap}
evolution equations. In this way, the origin of the differences of partons
densities in nucleons with respect to
nuclei is not addressed, but contained in the
parametrization at $Q_0^2$ which is obtained from a fit to experimental data.

The results from different models usually depend on additional
semiphenomenological assumptions and often contradict each other. For example,
concerning
the $Q^2$-dependence of the effect,
in \cite{nikolaev,nikolaev2,kope} it is argued that $q\bar q$
configurations of a large dimension give the dominant contribution to the
absorption, which results essentially independent of $Q^2$ (this is the case in
\cite{noso} until extremely small $x$, where a dependence $\propto
\ln{Q}/Q$ appear
related with the use of BFKL evolution \cite{bfkl}). On the other
hand, in the gluon recombination approach of \cite{mq} the absorption is
obtained as a clear higher-twist effect dying out at large $Q^2$. Finally,
in the models \cite{gnucl} which use DGLAP, all $Q^2$-dependence comes from
QCD evolution and is thus of a logarithmic, leading-twist
nature, see \cite{fgms} for a
comparison between multiple scattering and DGLAP approaches.
Predictions (particularly for the gluon density) on the $x$-evolution towards
small $x$ turn out to be very different \cite{salmor,amigo}.

In practice these studies are of uttermost importance to compute particle
production in collisions involving nuclei.
For example, in the framework of collinear
factorization \cite{facto} parton densities in the nucleus following
the spirit of \cite{gnucl} are needed, see e.g.
\cite{ekrt} for recent applications to
inclusive particle production in heavy ion collisions.
While this scheme is suitable to compute particle
production at scales $\Lambda_{QCD}^2\ll
Q^2\ \lsim \ s$ and thus in the hard region, for semihard
(minijet) production, $\Lambda_{QCD}^2\ll Q^2 \ll s$,
the $k_T$-factorization scheme \cite{ktfact} should become the suitable one
(in \cite{forhic,dima,pirner} applications to heavy ion collisions can be
found). Here the
tool that is needed is the so-called unintegrated gluon distribution (see a
precise definition below and \cite{smallx} for a review of its situation for
nucleons). Let us stress that different approaches give very
different predictions
for multiplicities at RHIC and LHC \cite{review}.

In this paper we present a simple model for nuclear structure functions in the
region of small $x$ ($\lsim \ 0.01$) and of small and moderate $Q^2$ ($\lsim \ 
20$
GeV$^2$) in the
dipole picture \cite{nikolaev,dipmue}. It consists on an extension to
nuclei,
using
the Glauber-Gribov picture \cite{glaugri}, of the saturation model for
the proton in
\cite{wust} without any new parameter
(in \cite{pirner} a similar strategy is used but with a simplified
dipole-nucleon cross section and no comparison with experimental data on
nuclear structure functions is performed; also in \cite{misha2} this ansatz
is used but just as an initial condition for a non-linear evolution
equation). The plan of the paper is as
follows: In the next Section the model will be described. In Section \ref{comp}
a comparison of the results of the model with available data on $F_{2A}$
will be shown. In Section
\ref{satur} the unintegrated gluon density obtained in this model, together
with the
saturation scale it implies, will be discussed. Finally, in the
last Section conclusions and possible applications of the model will be
outlined.

\section{Description of the model} \label{model}

The nuclear structure function $F_2$ can be standardly defined via
the cross sections $\sigma_{T,L}$  for the collision of
the transversal ($T$) or longitudinal ($L$) virtual
photon of momentum $q$, $q^2=-Q^2$, on the nucleus $A$
of momentum $Ap$:
\beq
F_{2A}(x,Q^2)=\frac{Q^2}{\pi e^2}(\sigma^A_T+\sigma^A_L); \label{eq1}
\eeq
this expression holds for the small $x$ region where we are going to work.
Both cross sections
can be conveniently presented in the dipole model \cite{nikolaev,dipmue}
via the cross section
$\sigma_{dA}(x,r)$,
for the scattering of a colour dipole of transverse
size $r$ on the nucleus:
\beq
\sigma^A_{T,L}(x)=\int d^2r\ \rho_{T,L}(r)\sigma_{dA}(x,r), \label{eq2}
\eeq
where $\rho_{T,L}$ are the distributions of colour dipoles
created by splitting of the incident photon into $q\bar q$ pairs
\cite{nikolaev}:
\beq
\rho_T(r)= \frac{e^{2}N_c}{8\pi^3}\sum_{f}z_f^2
\int_{0}^{1}d\alpha\ 
\Big\{[\alpha^{2}+(1-\alpha)^{2}]\epsilon^{2}{\rm K}_{1}^{2}(\epsilon r)+
m_f^2{\rm K}_0^2(\epsilon r)\Big\} \label{eq3}
\eeq
and
\beq
\rho_{L}(r)= \frac{e^{2}N_c}{2\pi^3}Q^2\sum_fz_f^2
\int_{0}^{1}d\alpha\ 
\alpha^2(1-\alpha)^2{\rm K}_{0}^{2}(\epsilon r). \label{eq4}
\eeq
Here summation goes over flavours (which will be limited to 3 or 4, see the
comments below Eq. (\ref{eq6})),
$\epsilon^2=Q^{2}\alpha (1-\alpha)+m_f^2$, and $m_f$ and $z_f$ are
respectively the mass and
electric charge in units of the proton charge
$e$, of the quark of flavour $f$.

For the total
cross section of a dipole on a proton we will use the saturation
model in
\cite{wust}, which provides a good description of inclusive and
diffractive experimental data on
$F_{2p}$ for
$x<0.01$ and small and moderate $Q^2< 20$ GeV$^2$. Other models which
also describe data in this region could be used, as that in \cite{model2}, see
\cite{cfks} for a comparison between this model and that of \cite{wust}, which
we will employ because of its simplicity. The form of the cross section
is\footnote{In this model no impact parameter of the proton is
explicitly given,
so it cannot be considered a pure eikonalization of some elementary
amplitude \protect{\cite{mueller2}};
on the contrary, in our extension to nuclei, Eqs.
(\protect{\ref{eq7}}) and (\protect{\ref{eq8}}) below, the impact parameter of
the nucleus is explicitly taken into account.}
\beq
\sigma_{dp}(x,r)=\sigma_0 \left[1-\exp{\left(-\frac{Q_s^2(x) r^2}{4}\right)}
\right],
\label{eq5}
\eeq
with
\beq
Q_s^2(x)=Q_0^2\left(\frac{x_0}{\tilde{x}}\right)^\lambda,\ \ \tilde{x}=x\left(
1+\frac{4m_f^2}{Q^2}\right).
\label{eq6}
\eeq
In \cite{wust}
some parameters were kept fixed: $Q_0^2=1$ GeV$^2$, $m_u=m_d=m_s=0.14$ GeV and
$m_c=1.5$ GeV. Three parameters were determined from a fit to data and
have values $\sigma_0=23.03$ (29.12) mb, $\lambda=0.288$ (0.277) and
$x_0=3.04\cdot 10^{-4}$ ($0.41\cdot 10^{-4}$) for the 3-flavour (4-flavour)
version of the model. $Q_s^2$ is called the saturation scale, which we will
discuss
in Section \ref{satur}.

Let us comment that, neglecting the influence of quark masses in
variable
$\epsilon$ which has \cite{geosca} small effect in the case of
light
quarks, this model implies, in its 3-flavour version, an exact scaling
of
the cross sections $\sigma^A_{T,L}$ with $\tau=Q^2/Q_s^2$. This scaling
has been shown \cite{geosca} to be fulfilled to a good approximation by
all DIS data for $x<0.01$ (which go up to $Q^2\sim 450$
GeV$^2$)\footnote{In the 4-flavour version, a flavour dependence is
introduced in $Q_s^2$ through variable $\tilde{x}$ and some deviation of
the scaling may also appear through variable $\epsilon$ in the photon
wave
function due to the larger charm mass.}.
So this model, which leads
for large $Q^2$ to Bjorken scaling, is apparently able to mimic along a wide
$Q^2$-region the QCD evolution (see \cite{ngbw} for improvements of this model
to include DGLAP evolution). This $\tau$-scaling has been argued to hold for
such a large $Q^2>Q_s^2$ ($\sim 1$ GeV$^2$ in the region where data are
available)
in the framework of semiclassical, high-density QCD models
\cite{cgcsca}, and has also been found in numerical solutions of the non-linear
evolution equations at small $x$ \cite{numer,misha}.
Let us also point that this model does not correspond to any fixed twist, see
\cite{twist} for a study of
its twist structure.

The extension of this model to the nuclear case can be made in a
straightforward manner in the Glauber-Gribov approach \cite{glaugri}:
ignoring isospin effects which are negligible at small $x$ where the
model will be applied, we will
substitute $\sigma_{dp}(x,r)$ by
\beq
\sigma_{dA}(x,r)=\int d^2b \ \sigma_{dA}(x,r,b),
\label{eq7}
\eeq
with $b$ the impact parameter of the center of the dipole
relative to the center of the nucleus and
\beq
\sigma_{dA}(x,r,b)=2\left[1-\exp{\left(-\frac{1}{2}AT_A(b)\sigma_{dp}(x,r)
\right)}\right]
\label{eq8}
\eeq
the total dipole-nucleus cross section for fixed impact parameter, with
$\sigma_{dp}(x,r)$ given by Eqs. (\ref{eq5}) and (\ref{eq6}). $T_A(b)$ is
the nuclear profile function (longitudinal integral of the nuclear density,
$T_A(b)=\int_{-\infty}^{\infty}dz\ \rho_A(z,\vec{b})$) normalized to unity,
$\int d^2b \  T_A(b)=1$; we employ a nuclear density in the form of a
3-parameter Fermi distribution with parameters taken from \cite{prof}.
With this normalization we recover the dipole-nucleon cross section
making a power expansion for small $AT_A(b)\sigma_0$, keeping the first term and
putting $A=1$.
Also the centrality (impact parameter)
dependence of the structure functions
can be computed by direct substitution of $\sigma_{dA}(x,r,b)$, Eq. (\ref{eq8}),
into Eq.
(\ref{eq2}). This model implies a new scaling for nucleus of the type of
the $\tau$-scaling for proton and a new saturation scale,
which we will discuss in Section \ref{satur}. These two Eqs. (\ref{eq7})
and (\ref{eq8}) constitute the central point in our extension of the model of
\cite{wust} to the nuclear case.

Using Eqs. (\ref{eq1})-(\ref{eq8}) we can compute the nucleon and nuclear
structure functions and the corresponding ratios, which we will compare with
experimental data in the next Section. But first let us discuss the region of
applicability
of the model: This should be that of small $x$ (due to the use of the model
of \cite{wust} for the nucleon,
to the neglection of isospin effects and to the use of (\ref{eq8}) which
requires a large coherence length, achieved at small $x$,
of the photon fluctuation),
and of small and
moderate $Q^2< 20$ GeV$^2$. Although the $\tau$-scaling may suggest that the
region of applicability in $Q^2$ could be wider, we think that
a safe extrapolation to higher $Q^2$ would require to
implement DGLAP evolution \cite{ngbw}.
Also the extrapolation to very small $x$ could
imply effects of gluon or pomeron fusion
like those included in the non-linear evolution equations
mentioned previously, so it should be taken with care. Nevertheless, numerical
studies \cite{misha} show that the onset of the non-linear effects is quite
smooth, becoming large for extremely small $x$. So we consider\footnote{In
\protect{\cite{misha}}
it is seen that the scaling induced by the non-linear evolution is
fulfilled for values of $y=[\pi/(N_c\alpha_s)]\ln{(x_0/x)}$ greater than
$2.2$, which for $\alpha_s=0.2$ and
$x_0=0.01$ means $x$ smaller than $10^{-7}$.} this model as a
reasonable approximation for values of $x\ \gsim \ 10^{-5}\div 10^{-6}$, which
are those relevant for RHIC and LHC.
Due to the fact that for nuclear
structure functions the amount of experimental
data is much more limited than that for nucleons,
we will perform the comparison with experimental data for $x<0.02$.

\section{Comparison with experimental data} \label{comp}

Here we show the results of the model together
with available experimental data. In
Figs. 1-5, the results of the model in the 3-flavour (solid
lines)
and 4-flavour (dashed lines) versions (see the explanations below Eq.
(\ref{eq6})) are compared with experimental data
from \cite{e665,nmc,nmcq2,oldnmc}. In these Figs.,
$R(A/B)=[BF_{2A}(x,Q^2)]/[AF_{2B}(x,Q^2)]$.
Even when joined with a line, the results
of the model have been computed at the same ($\langle x\rangle,
\langle Q^2\rangle$) as the experimental data; in the latter,
the inner error bars
are the statistical errors, while the outer error bars show statistical and
systematic errors added in quadrature (with overall normalization
uncertainties ignored). Considering both
the absence of any free parameter to go from the
nucleon to the nuclear case and the simplicity of the model, we find the
agreement quite reasonable. Concretely, from Figs. 3 and 4 it
can be concluded that the $A$- and $Q^2$-dependences of the data at fixed $x$
are well reproduced, while in Figs. 1, 2 and 5 the
$x$-dependence is reasonably reproduced, taking into account that these latter
Figures
contain data with $Q^2$ going from $\sim 0.15$ GeV$^2$ for the smallest $x$
values to $\sim 4.5$ GeV$^2$ for the highest $x$.

As the version with 3 flavours is simpler, gives an equally
reasonable agreement with nuclear data than the 4-flavour version, and produces
a better description of the nucleon data \cite{wust},
from now on we will restrict our computations to the 3-flavour version.

Let us turn now to the behaviour of the ratio of longitudinal to transverse
cross sections. Experimentally \cite{hermes} large nuclear effects have been
observed: the ratio $\sigma_L/\sigma_T$ in N ($^3$He) over $\sigma_L/\sigma_T$
in D has been found to reach values as high as $\sim 5$ ($\sim 2$)
for $0.01<x<0.03$ and
$Q^2<1$ GeV$^2$. Some explanations \cite{ht} point to nuclear enhanced power
corrections, but the experimental data are under reanalysis
\cite{newhermes} and the evidence
of such strong
nuclear effects is now dubious. While this important point has to be settled, it
is clear than in our model such strong effects are not present, as the nuclear
effects are contained in the Glauber-Gribov cross section in Eqs. (\ref{eq7})
and (\ref{eq8}) which is common to both longitudinal and transverse
cross sections, see Eq. (\ref{eq2}). In Fig. 6 we show the results in our model
for the ratio $\sigma_L/\sigma_T$ in nucleus over $\sigma_L/\sigma_T$ in proton,
for C and Pb.
It can be seen that the nuclear effects never go beyond $\pm 12$ \%, which is a
clear prediction of our model.

In Fig. 7 we present predictions of the model for the ratio $F_{2A}/(AF_{2p})$
for C and Pb, together with the $x$-evolution for Be, C, Al, Ca, Fe, Sn and Pb
at fixed $Q^2=2.25$
GeV$^2$. A clear evolution with $Q^2$ can be seen, which in this model is due
to the interplay between the (transversal and longitudinal)
probabilities to get a dipole of size $r$, Eqs. (\ref{eq3}) and (\ref{eq4}), and
the dipole-target cross section, Eqs. (\ref{eq5}), (\ref{eq7})
and (\ref{eq8}), and
cannot be addressed to any concrete twist but to an admixture
of all twists, see \cite{twist}. At large
enough $Q^2$ this dependence on $Q^2$ will eventually disappear, as this
model, as stated previously, leads to Bjorken scaling.

As a last point in this Section, let us comment on other possible options to
get nuclear structure functions in the framework of the dipole
model.
A
simple form for the dipole-nucleus cross section is suggested by high-density
QCD \cite{semi,mueller1,mueller2}:
\beq
\sigma_{dA}(x,r)=\int d^2b \ 
\left[1-\exp{\left(-\frac{Q^2_{sA}(b) r^2}{4}
\right)}\right].
\label{eq9}
\eeq
We have tried several relations between $Q^2_{sA}$ (the saturation scale in
nuclei) and that in proton, $Q^2_s$. On the one hand, we have used a
relation coming from the running of the coupling, of the type
\beq
Q^2_{sA}
\ln{\left(\frac{Q^2_{sA}}{\Lambda_{QCD}^2}\right)}\propto
\left(\frac{T_A(b)}{T_A(0)}\right) A^{1/3} Q^2_{s}
\ln{\left(\frac{Q^2_{s}}{\Lambda_{QCD}^2}\right)}.
\label{eq10}
\eeq
On the other hand, we have imposed the
first scattering approximation (valid for $r\to 0$) in Eq. (\ref{eq8}),
\beq
Q^2_{sA}=\frac{1}{2}AT_A(b)\sigma_0 Q^2_{s}
\label{eq11}
\eeq
(in this expression the value of the running coupling evaluated at the
appropriate scale is hidden in $\sigma_0$, see e.g. \cite{mueller2})\footnote
{The fact that both $Q^2_{sA}$ and $Q^2_{s}$ may have roughly the
same $x$-dependence can be justified by the following qualitative argument:
$Q^2_{sA}$ is related with
$p_T$-broadening in nucleus, $Q^2_{sA}=n_A(b) Q^2_{s}$, with $n_A(b)$ the number
of scatterings at impact parameter $b$. As $n_A(b) \propto AT_A(b) xG
\alpha_s(Q_{sA}^2) /Q_{sA}^2$ (for a perturbative
QCD cross section evaluated at scale
$Q_{sA}^2$ and $xG$ the gluon distribution in a nucleon)
and $Q^2_{s}\propto xG$, both $Q_{sA}^2$ and $Q^2_{s}$ show
the same $x$-behaviour (modulo the logarithm coming from the running coupling).
Special thanks are given to D. E. Kharzeev for discussions and
suggestions on all these points.}.
But so far we have not succeeded in getting a satisfactory description of 
experimental data, not even on a qualitative level: either too strong a
shadowing is observed or too fast an evolution in $x$ (and too slow in $Q^2$)
is obtained.
Indeed Eqs. (\ref{eq8}) and (\ref{eq9}) contain different physical
assumptions on the nature of the scattering centers: while (\ref{eq8})
considers multiple scattering on single nucleons (described by the saturating
form (\ref{eq5})), (\ref{eq9}) implies scattering on a black area filled
with partons coming from many nucleons.
Our lack of success in reproducing the experimental data with (\ref{eq9})
suggests that higher order rescatterings are actually
needed in the exponent of
Eq. (\ref{eq5}) for the proton,
and that the asymptotic region where Eq. (\ref{eq9})
should be valid to describe data on $F_{2A}$ integrated over impact parameter,
is not reached yet (i.e. the grey region is still dominating the
scattering); a very important test for the form (\ref{eq9}) would be its
ability to describe experimental data on diffraction (which is indeed fulfilled
by (\ref{eq5})
\cite{wust}).
In the next section we will address the behaviour of the
saturation scale in our model.

\section{Unintegrated gluon distribution and saturation scale} \label{satur}

As stated in the Introduction, in the $k_T$-factorization scheme
\cite{ktfact} a key ingredient
is the unintegrated gluon distribution of the hadron,
$\varphi_A(x,k,b)$ (sometimes it appears in the literature as $f=k^2\varphi$
\cite{forhic,smallx}), with $k$ the transverse momentum.
This $\varphi_A(x,k,b)$ at fixed impact parameter $b$
is related, at lowest order in $k_T$-factorization \cite{smallx},
to the dipole-nucleus cross section by a
Bessel-Fourier transform (see \cite{misha,misha2}):
\beq
\varphi_A(x,k,b)=-\frac{N_c}{4\pi^2\alpha_s}k^2 \int \frac{d^2r}{2\pi}
\ \exp{(i\vec{k}\cdot \vec{r})}
\sigma_{dA}(x,r,b),
\label{eq12}
\eeq
with $k^2=\vec{k}\cdot \vec{k}$, $r^2=\vec{r}\cdot \vec{r}$ and vectors
defined
in the two-dimensional transverse space.
The unintegrated gluon can be related to the `ordinary' gluon density (that used
in collinear factorization \cite{facto}) by
\beq
xG(x,Q^2,b)=\int^{Q^2} dk^2\  \varphi_A(x,k,b),
\label{eq13}
\eeq
although this expression must be considered with great care, as it is only true
for large $Q^2\gg Q_s^2$ (the actual relation is not with the collinear glue but
with the gluon distribution in the light-cone wave function of the hadron, see
\cite{semi,mueller1,mueller2}).

For the proton, Eq. (\ref{eq12})
leads to the result $\varphi_p(x,k)
\propto \sigma_0(k^2/Q_s^2)
\exp{(-k^2/Q_s^2)}$ \cite{wust}\footnote{In some proposals
\cite{mueller1,dima}
it is considered that the unintegrated gluon distribution should
tend to a constant as $k\to 0$. For discussions on the `correct' definition
and behaviour of this quantity, see \cite{misha,pirner}.}.
For the nucleus, using the technique outlined
in the Appendix of \cite{misha2} (or simply applying $[N_c/(2\pi^2
\alpha_s)]k^2\nabla_k^2$ to function
$\phi_0$ defined in Eqs. (31) and (34) in that Reference), we get
\beq
\varphi_A(x,k,b)=\frac{N_c}{\pi^2\alpha_s}
\frac{k^2}{Q_s^2}\sum_{n=1}^\infty \frac{(-B)^n}
{n!} \sum_{l=0}^n C_n^l \frac{(-1)^l}{l} \exp{\left(-\frac{k^2}{lQ_s^2}\right)},
\label{eq14}
\eeq
with
\beq
B=\frac{1}{2}AT_A(b)\sigma_0.
\label{eq15}
\eeq
As in the case of proton, Eq. (\ref{eq14}) shows explicit scaling in
$k^2/Q_s^2$; besides, in this Equation
(as in (\ref{eq8})) the result for proton is recovered
making a power expansion for small $B$, keeping the first term and
putting there $A=1$.
For realistic values of $B<3$, Eq. (\ref{eq14}) turns out to
be very suitable for numerical computations, as the convergence of the series
in $n$ is very fast and only a few terms are needed to get
$\varphi_A(x,k,b)$ to the desired accuracy.

In Fig. 8 we show the unintegrated gluon distribution for proton, and for Pb
in three cases: central ($b=0$), peripheral ($b=7$ fm), and integrated over
$b$, and for two values of $x=10^{-2}$ and $10^{-6}$ (in these computations
there is no $Q^2$, so the substitution in Eq. (\ref{eq6}) is meaningless here
and we will make no distinction in this Section between $x$ and $\tilde{x}$;
this also avoids the complication of flavour dependence in case the 4-flavour
version is to be used, although as stated previously we will use the
3-flavour version). A scaling in $k^2/Q_{sA}^2$ (with $Q_{sA}^2$
identified with the position of the maximum, see below) is
perfectly visible, as in each case
the curves move to the right with decreasing $x$
while keeping their shape and size (this phenomenon has also been
found in the framework of the
non-linear equations for small $x$
\cite{semi,mueller1,numer,misha} and the solution
called a soliton wave \cite{misha}). Besides it can be seen that the shape
of the curves for different cases is quite close, the only differences
being the height, the position of the maximum which at fixed $x$ varies
from left to
right for proton and Pb with increasing centrality, and the logarithmic
width which slightly increases
with increasing centrality (being e.g. at 1/10 the maximum height,
2.20 for proton and 2.40 for central Pb).

Keeping in mind the
difficulties to identify at small and moderate $Q^2$
the integral of the unintegrated gluon
distribution with the ordinary gluon density, Eq. (\ref{eq13}) (see the
comments below it),
it is still tempting to use this Equation and compare with other
approaches. A comparison at $Q^2=5$ GeV$^2$
of the results
of our model with
others, for the ratio of gluon densities in Pb over proton, can be found in
\cite{amigo}. There it can be seen that our results at $x\simeq 10^{-2}$
roughly coincide with those of \cite{gnucl,ina} and are higher than
those of
\cite{fgms,newhijing}, while at $x\simeq 10^{-5}$ they become smaller
than those of \cite{gnucl,ina}, get close to those of \cite{fgms} and
are still larger than those of \cite{newhijing}. Apart from the
constraints coming from existing DIS experimental data on nuclei which are very
loose for the glue at small $x$, in \cite{gnucl}
the saturation of gluon shadowing comes mainly
from the initial condition for DGLAP evolution where
this saturation has been imposed, while \cite{fgms,ina} are theoretical models
and in \cite{newhijing} the behaviour of the glue has been fixed in
order to reproduce charged particle multiplicities in AuAu collisions
at RHIC.
Additional caution has to be taken to compare our results with those
coming from DGLAP analysis \cite{gnucl}: our ratios for the glue at some
moderate, fixed $Q^2$ and very small $x$
result smaller than the ratios for $F_2$ at the same $x$, $Q^2$, which leads to
problems with leading-twist DGLAP evolution \cite{salmor};
as our model mimics the DGLAP increase of $F_2$ along a
wide
$Q^2$-region,  this
is most probably related with the use of Eq. (\ref{eq13})
at too small $Q^2$.
In any case,
our model leads to Bjorken scaling at some $Q^2$ which increases
with decreasing $x$:
the $Q^2$-independence of the gluon density is achieved in our model 
at $Q^2\simeq 15$ GeV$^2$ for $x=10^{-2}$ and at $Q^2\simeq 300$ GeV$^2$
for $x=10^{-6}$. Of course this is due to the exponential decrease of
the unintegrated gluon at large $k^2$, see (\ref{eq14}); other proposals
(see e.g. \cite{mueller1,dima}) consider a
behaviour $\propto 1/k^2$ at large $k^2$ which obviously leads to a
logarithmic, DGLAP-like increase of the gluon density with increasing
$Q^2$.

Now we turn to the
saturation scale. While for the case of the proton its definition is quite
clear in coordinate space, for the nucleus a clean definition is better
obtained in momentum space, where it corresponds to the maximum of
the unintegrated gluon distribution \cite{mueller1,numer,misha,newmueller}.
The results are shown in Fig. 9 (upper plot) for the same cases as in Fig. 8.
Contrary to naive expectations, the saturation scale between proton and
central nucleus is not simply proportional to $A^{1/3}$ but has a prefactor
which makes the result smaller (turning the expected factor 5.9 for Pb into
a factor 2.0). This can be understood using analytical estimates. Taking the
exponent in Eq. (\ref{eq8}) to be 1/4 (in analogy to the case of the proton), 
we get
\beq
Q_{sA}^2\simeq
\left[4\ln{\left(\frac{2AT_A(b)\sigma_0}{2AT_A(b)\sigma_0-1}\right)}
\right]^{-1} Q_s^2,
\label{eq16}
\eeq
which gives a $Q_{sA}^2\simeq 2.4 Q_s^2$ for Pb at $b=0$. For $AT_A(b)\sigma_0
\gg 1$, we get exactly Eq. (\ref{eq11}), recover the expected
proportionality $Q_{sA}^2\propto A^{1/3} Q_s^2$ and find
with this asymptotic form $Q_{sA}^2\simeq 2.5 Q_s^2$. As
the form (\ref{eq8}) gives a nice agreement with data and (\ref{eq9}),
(\ref{eq11}) do not, we conclude that preasymptotic contributions (i.e.
not corresponding to $AT_A(b)\sigma_0 \to \infty$) play
a dominant role for the description of present available data.

To perform some comparison
with other approaches, let us see the saturation scale coming
from the numerical solution of the non-linear
evolution equation in \cite{misha}. There a form (for the evolution obtained
starting from a phenomenologically motivated initial distribution) is
obtained that can be approximated, for central ($b=0$) Pb
by the equation
\beq
Q_{sA}^2 \simeq
A^{0.37}\exp{\left[-6.43+0.78\ln{\left(x_0 \over x\right)}\right]};
\label{eq17}
\eeq
here $x_0=0.01$, a value of $\alpha_s=0.2$ has been used and the result is
in GeV$^2$. The coefficients in this Equation coincide with those obtained in
other numerical and analytical studies \cite{mueller1,numer,cgcsca}. Its
results for $A=1$ and $A=208$ are shown in Fig. 9 (lower
plot), but one should keep in mind the way in which
the coefficients in Eq. (\ref{eq17})
were obtained \cite{misha}:
They come from a fit to the position of the maximum of
the unintegrated
gluon distributions for different centralities and values of $x$
but when
the scaling induced by the non-linear evolution has already
set in. So, from
the first footnote it can be concluded that they correspond to $x<10^{-7}$ and
thus to a much lower value of $x$ than the region where we have actually
plotted them
in Fig. 9. From the comparison of the two plots in Fig. 9 it is clear that
the non-linear evolution produces a much steeper behaviour compared with
the model in this paper (i.e. an exponent $0.78$ compared to $0.288$),
while the asymptotic $A$-dependence is not really different ($0.37\simeq 1/3$).

\section{Conclusions} \label{conclu}

A simple model for nuclear structure functions in the region of small $x$,
and small and moderate $Q^2$, has been presented. It is a
parameter-free extension of the saturation model for the nucleon of \cite{wust}
in the
Glauber-Gribov approach, although in principle any saturating model which
correctly describes nucleon data, e.g. \cite{model2}, could be used.
This simple extension of the model for the nucleus should be valid until
non-linear evolution effects \cite{semi,balit,mueller1,numer,levin,misha,misha2}
become of importance, and for not too high $Q^2$ where
implementation of DGLAP evolution should be required. Conservative
estimates establish the region
of validity of the model to be $10^{-5}\div 10^{-6} < x < 10^{-2}$ and
$Q^2< 20$ GeV$^2$.

A reasonable agreement with experimental data is
obtained for the $x$-, $A$- and $Q^2$-dependence. The
longitudinal-to-transverse
cross section ratios show nuclear effects smaller than $\pm 12$ \%.
Predictions of the model for values of $x$ smaller than those
available to present experiments have been given.

The unintegrated gluon
distribution coming from our dipole-nucleus cross section
has been obtained and studied. Besides the behaviour of the saturation scale
in this model has been shown and discussed. It turns out to behave
$\propto A^{1/3}$ for large $A$ as expected, but large prefactors make the
resulting scale for nuclei smaller than naive estimates, e.g. for central Pb
2 times bigger, instead of 6 times, than the corresponding scale for proton.

Our model could be used to provide the starting condition for DGLAP evolution,
as performed in the approaches of \cite{gnucl}, for some
initial scale $Q_0^2\gg \Lambda_{QCD}^2$. The unintegrated gluon
distribution could be employed to compute particle production, using the
$k_T$-factorization scheme \cite{ktfact,smallx}, in high energy collisions
involving nuclei \cite{forhic,dima,pirner}. Work along these directions is in
progress.

As a last comment, our study implies the
existence of a
scaling for nuclei of the same type as that for proton \cite{geosca}, with
a non-trivial relation for the
$A$-dependence of the saturation scales between both cases,
while the $x$-dependence turns out to be the same. An
experimental, model-independent
extraction of the saturation scale in nuclei would be very
useful to settle the discussions on the relevance of the semiclassical
approach for existing or future experiments, and the region of validity
of perturbative QCD and DGLAP evolution.
Such an issue would be best explored,
and our model tested,
in high
energy lepton-nucleus colliders \cite{future}.

\vskip 1cm

\noi {\bf Acknowledgments:}
The author expresses his gratitude to M. A. Braun,
R. Engel, K. Itakura,
V. J. Kolhinen, L. McLerran, C. Pajares, P. V. Ruuskanen,
R. Venugopalan and U. A. Wiedemann, for useful discussions. Special thanks are
given to K. J. Eskola and
D. E. Kharzeev
for their interest, suggestions and encouragement to write this article, and
to A. Capella, E. G. Ferreiro and C. A. Salgado for discussions, suggestions and
a critical reading of the manuscript.
He also thanks CERN Theory Division, Departamento de F\'{\i}sica de
Part\'{\i}culas at Universidade de Santiago de Compostela, Department of
Physics at University of Jyv\"askyl\"a, Helsinki Institute of
Physics and Physics Department at BNL, for
kind hospitality during stays in which parts of this work have been developed.
Finally he acknowledges financial
support by CICYT of Spain under contract
AEN99-0589-C02 and by Universidad de C\'ordoba.

\section*{Figure captions:}
 
\noi {\bf Fig. 1.} 
Comparison of the results of the model (see text)
in the 3-flavour (solid lines) and
4-flavour (dashed line) versions with experimental data at small $x$ for
the ratios C, Ca and Pb over D, from
\cite{e665}.

\noi {\bf Fig. 2.} Id. to Fig. 1 but with data for Be, Al, Ca, Fe, Sn and Pb
over C, from \cite{nmc}.

\noi {\bf Fig. 3.} Comparison of the $A$-evolution
of the results of the model
in the 3-flavour (open circles) and
4-flavour (open triangles) versions with experimental data at small fixed $x$
for Be, Al, Ca, Fe, Sn and Pb over C, from
\cite{nmc}.

\noi {\bf Fig. 4.} Comparison of the $Q^2$-evolution
of the results of the model
in the 3-flavour (solid lines) and
4-flavour (dashed line) versions with experimental data for Sn over C
at small fixed $x$, from
\cite{nmcq2}.

\noi {\bf Fig. 5.} Id. to Fig. 1 but with data for C and Ca over D,
from \cite{oldnmc}.

\noi {\bf Fig. 6.} Results of the model for the ratio 
$\sigma_L/\sigma_T$ in C and Pb over $\sigma_L/\sigma_T$ in proton versus $x$.
In the plots, lines going from the bottom to the top correspond to
$Q^2=0.1$, 0.5, 1, 2.25, 5, 10 and 100 GeV$^2$.

\noi {\bf Fig. 7.} Results of the model for the $x$-dependence of
$F_{2A}/(AF_{2p})$. In the two upper plots, results for C (upper plot)
and Pb (plot
at the middle) versus $x$ are given for $Q^2=0.1$, 0.5, 1, 2.25, 5, 10
and 100 GeV$^2$ (lines going from the bottom to the top).
In the plot at the bottom,
$F_{2A}/(AF_{2p})$ is drawn versus $x$ at $Q^2=2.25$ GeV$^2$ for Be,
C, Al, Ca, Fe, Sn and Pb (lines going from the top to the bottom).

\noi {\bf Fig. 8.} Upper plot: 
Results of the model (in GeV$^{-2}$) for the
unintegrated gluon distribution for proton (solid line)
and for Pb integrated over $b$ (dashed line).
Lower plot: Results of the model for the
unintegrated gluon distribution for peripheral ($b=7$ fm, solid line)
and central ($b=0$, dashed line) Pb. In each case, two curves are
provided
for two values of $x=10^{-2}$ (curve to the left) and $10^{-6}$
(curve to the right).

\noi {\bf Fig. 9.} 
Upper plot: saturation momentum in the model for proton (solid line),
and for Pb
in three cases: central ($b=0$, dashed-dotted line),
peripheral ($b=7$ fm, dashed line), and integrated over
$b$ (dotted line). Lower plot: saturation momentum in the
numerical solution of the non-linear equation in \cite{misha},
Eq. (\ref{eq17}), for
$A=1$ (solid line) and $A=208$ (dashed-dotted line).
Notice the difference in vertical scales between the plots.

\newpage
\centerline{\bf \Large Figures:}
 
\vskip 3cm
 
\begin{center}
\epsfig{file=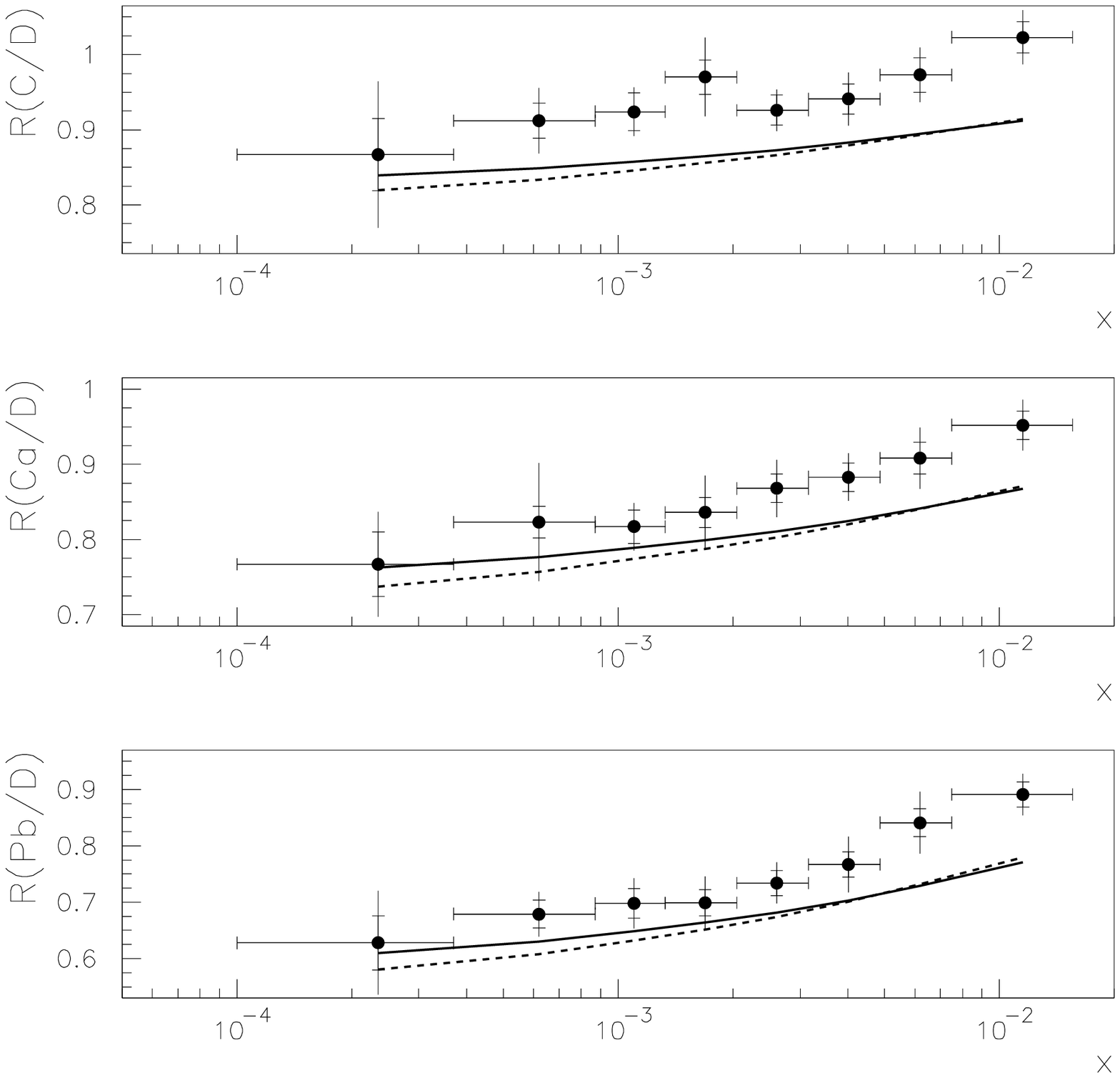,width=15.5cm}
\vskip 1cm
{\bf \large Fig. 1}
\end{center}
 
\newpage
 
\begin{center}
\epsfig{file=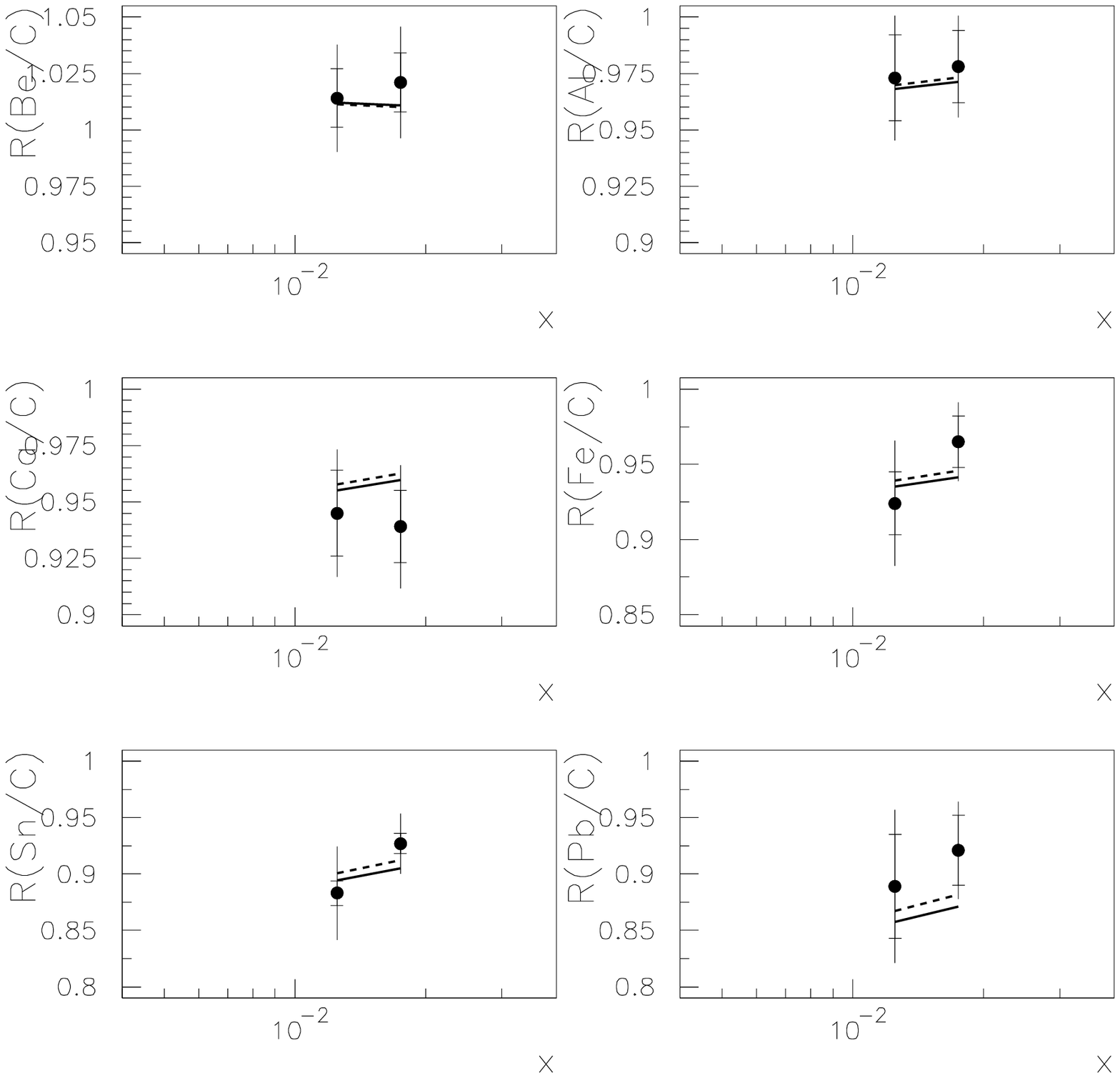,width=15.5cm}
\vskip 1cm
{\bf \large Fig. 2}
\end{center}
 
\newpage
 
\begin{center}
\epsfig{file=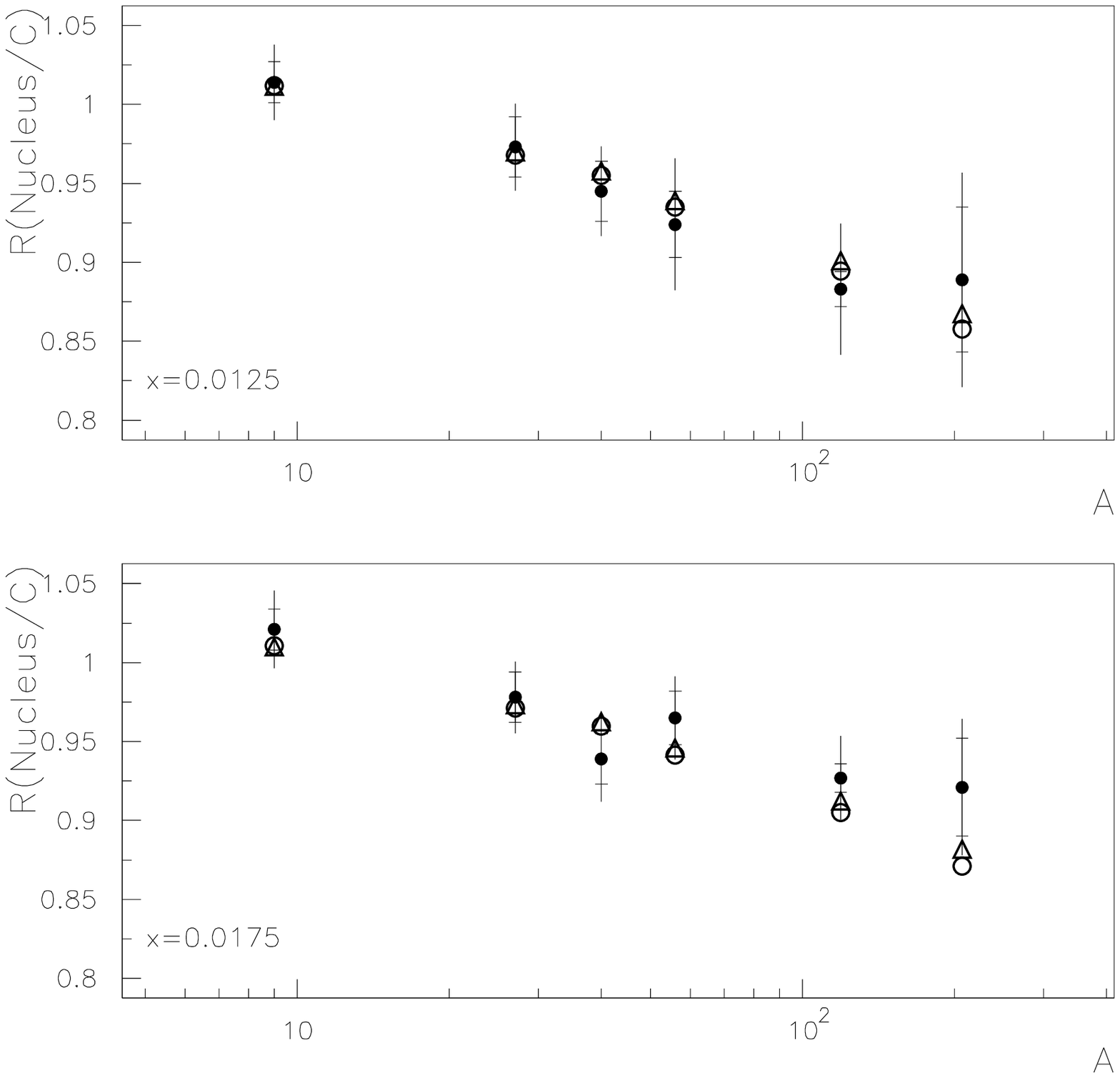,width=15.5cm}
\vskip 1cm
{\bf \large Fig. 3}
\end{center}
 
\newpage
 
\begin{center}
\epsfig{file=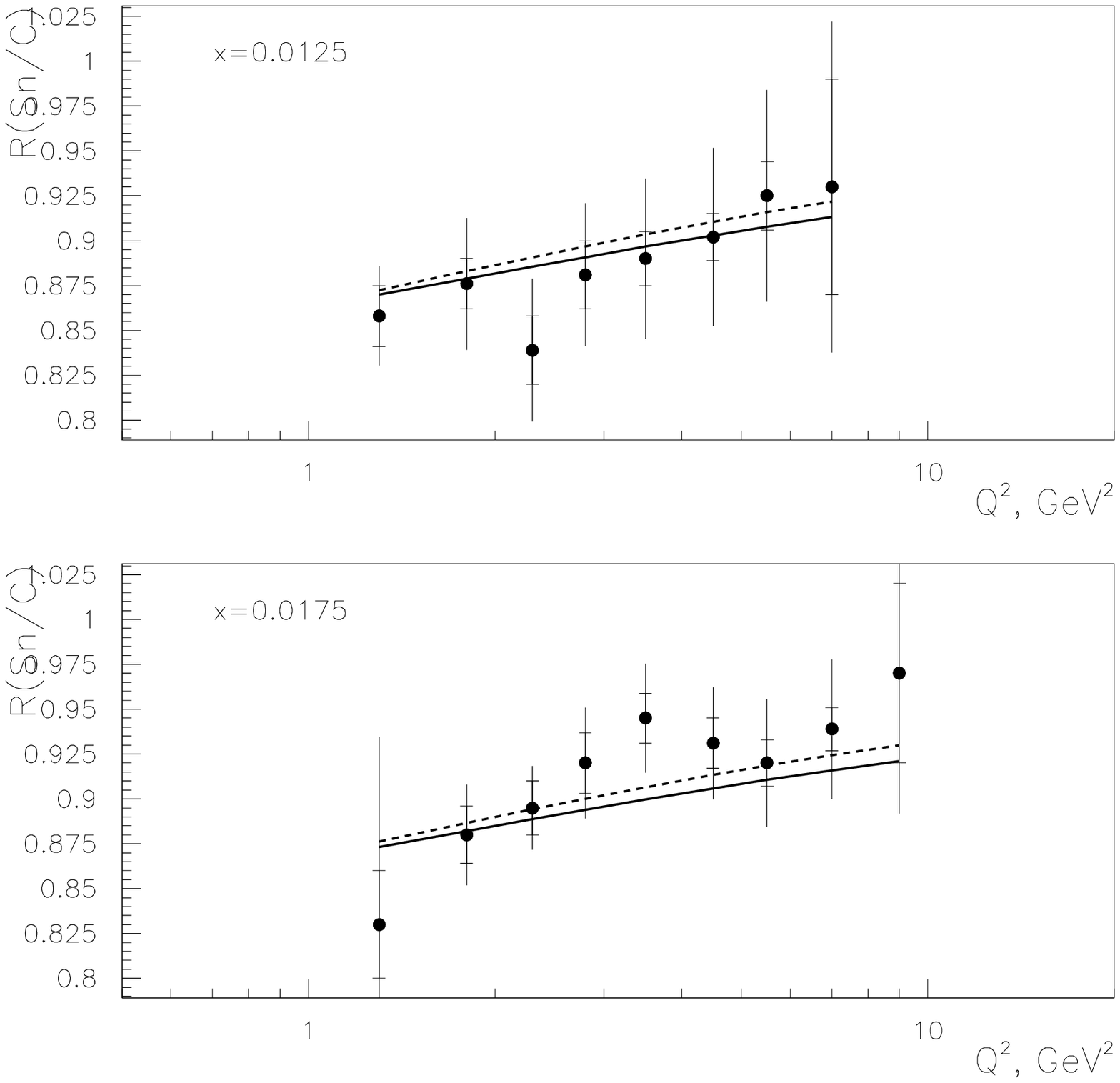,width=15.5cm}
\vskip 1cm
{\bf \large Fig. 4}
\end{center}
 
\newpage
 
\begin{center}
\epsfig{file=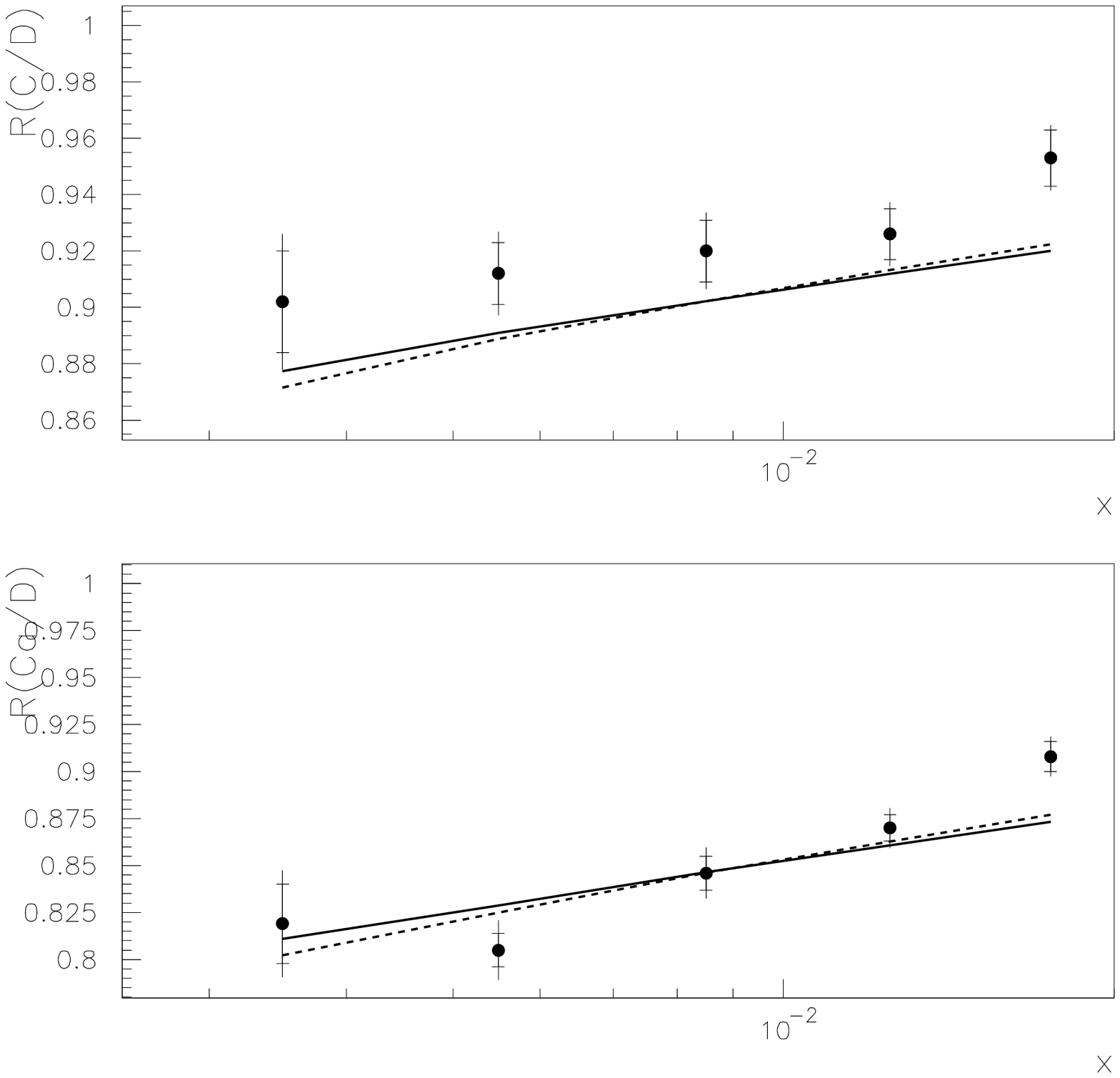,width=15.5cm}
\vskip 1cm
{\bf \large Fig. 5}
\end{center}
 
\newpage
 
\begin{center}
\epsfig{file=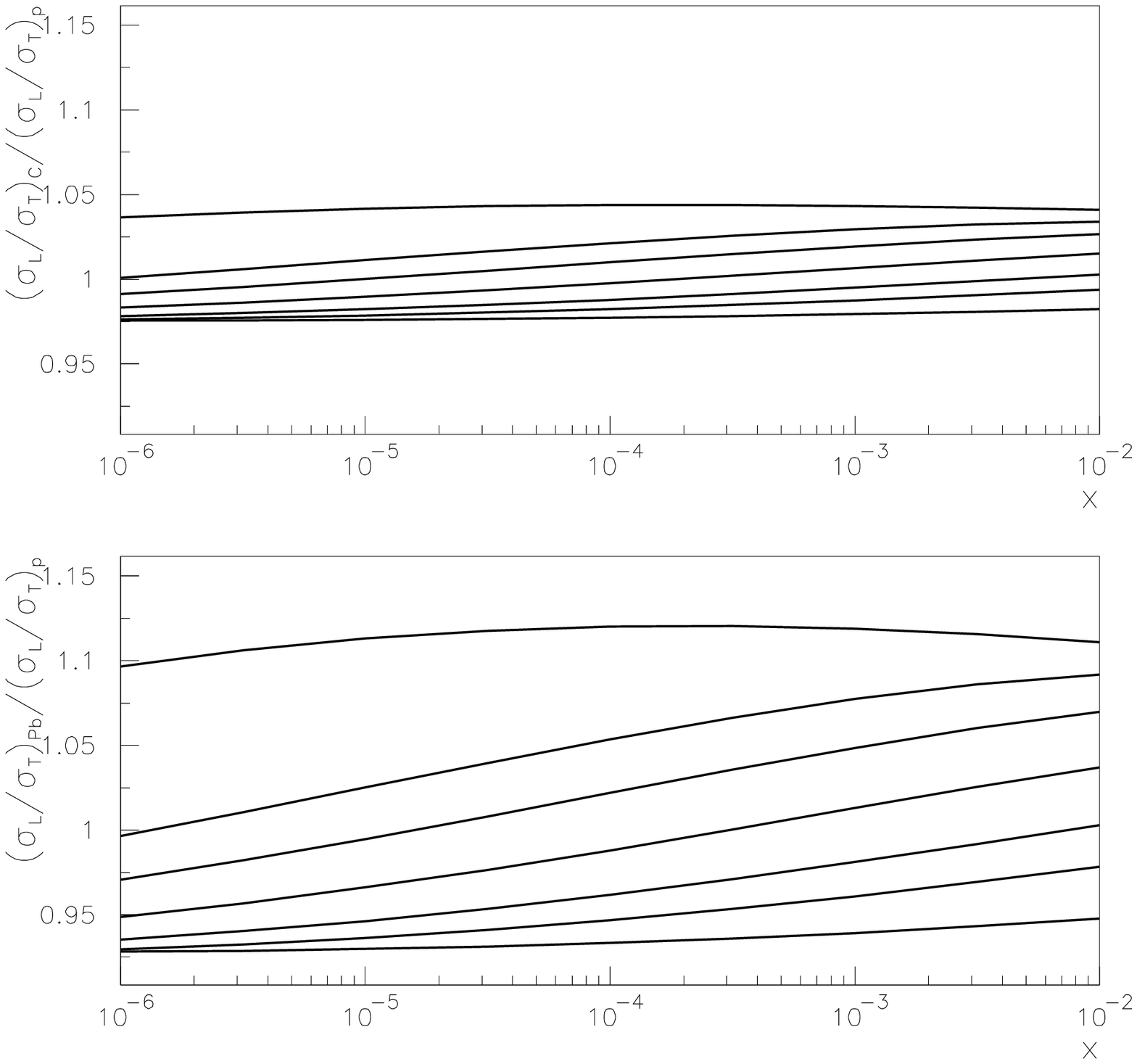,width=15.5cm}
\vskip 1cm
{\bf \large Fig. 6}
\end{center}
 
\newpage
 
\begin{center}
\epsfig{file=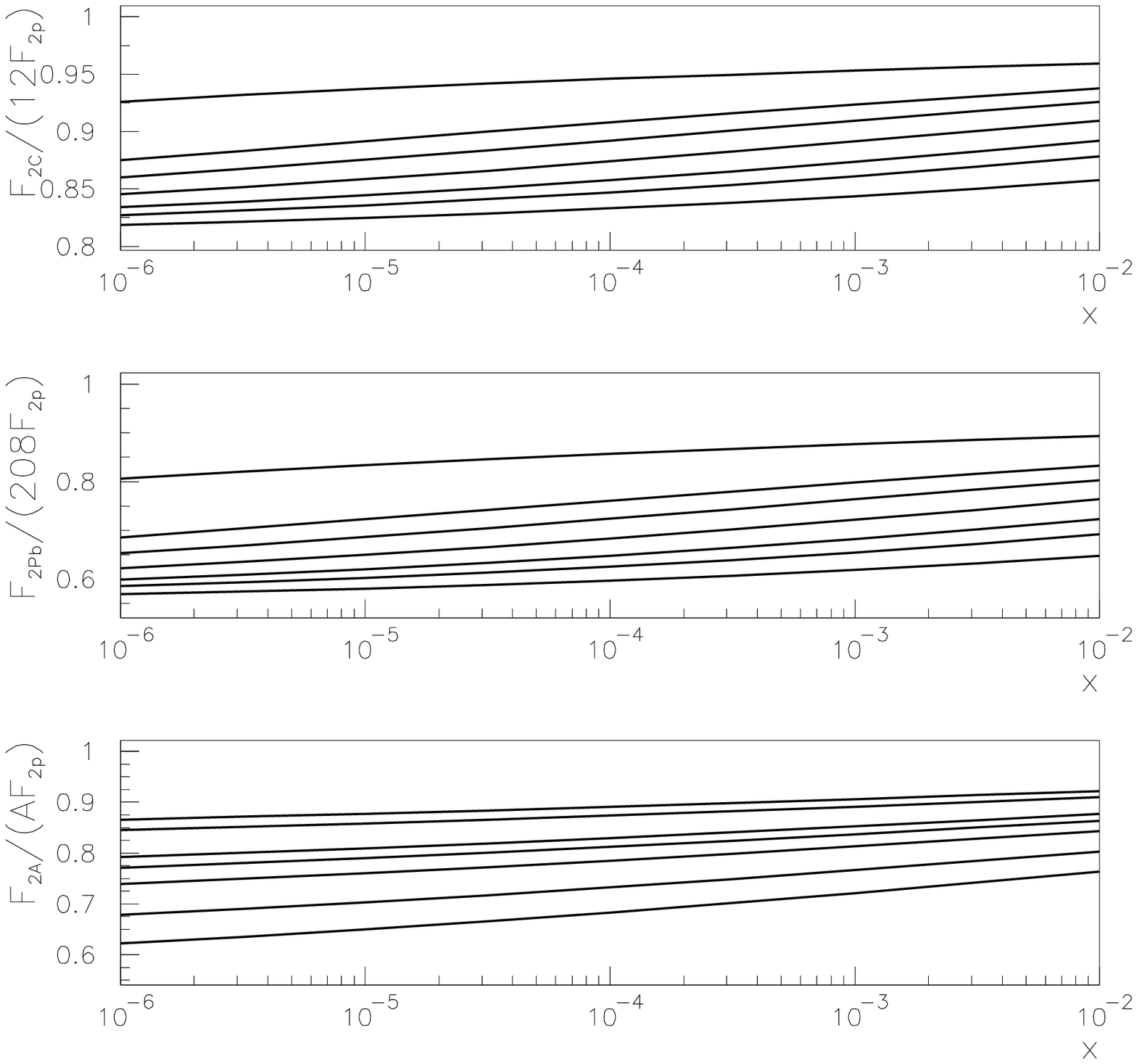,width=15.5cm}
\vskip 1cm
{\bf \large Fig. 7}
\end{center}
 
\newpage
 
\begin{center}
\epsfig{file=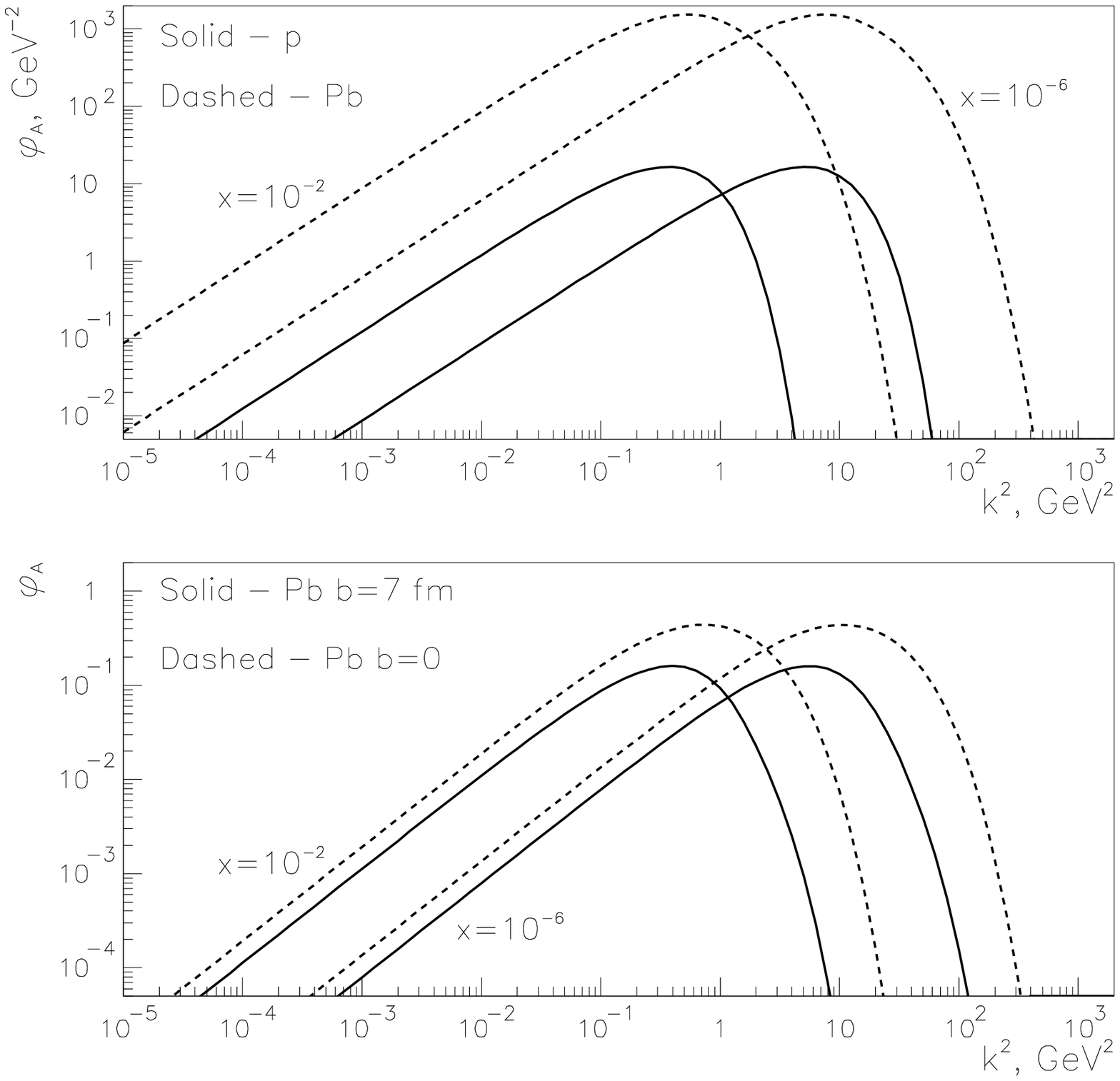,width=15.5cm}
\vskip 1cm
{\bf \large Fig. 8}
\end{center}

\newpage
 
\begin{center}
\epsfig{file=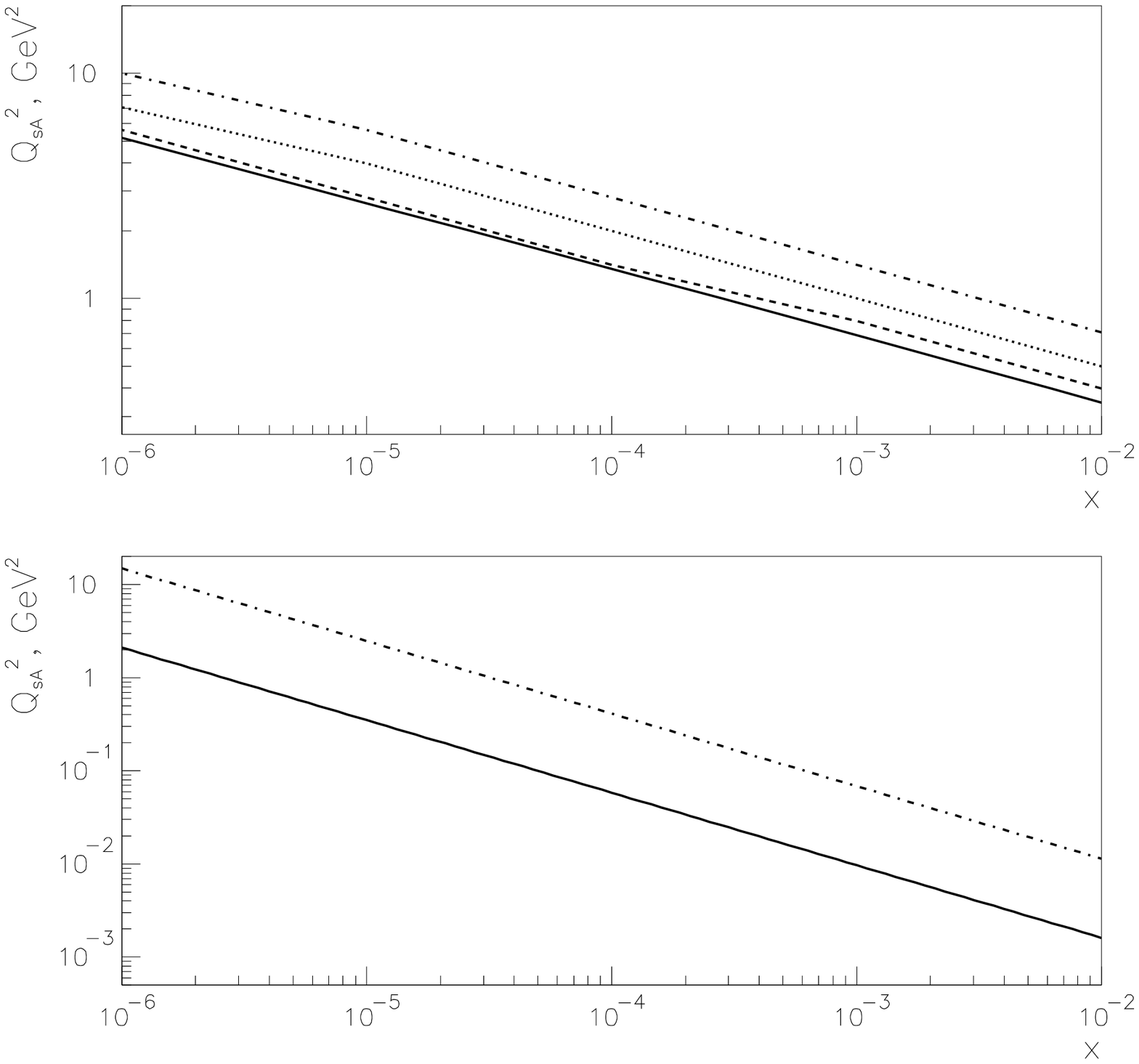,width=15.5cm}
\vskip 1cm
{\bf \large Fig. 9}
\end{center}

\end{document}